\documentclass[aps,prl,twocolumn,superscriptaddress,longbibliography,floatfix]{revtex4-2}
\usepackage{graphicx}
\usepackage{amsmath,amssymb,bm}
\usepackage{dcolumn}
\usepackage{color}
\usepackage[normalem]{ulem} 
\let\isout\sout \renewcommand{\sout}[1]{\ifmmode\text{\isout{\ensuremath{#1}}}\else\isout{#1}\fi}
\usepackage[hidelinks]{hyperref}
\hypersetup{pdftitle={Boundary-pinned nonsingular topologically charged defects in nanoplasmonics},
 pdfauthor={Michael I. Tribelsky and Vladimir V. Ternovski},
 pdfsubject={Letter, nanoplasmonics},
 pdfkeywords={nanoplasmonics, Poynting vector, topological defects, singular optics}}
\newcommand{\sgn}{\operatorname{sgn}}
\newcommand{\ii}{\mathrm{i}}
\newcommand{\ee}{\mathrm{e}}

\newcommand{\eps}{\varepsilon}

\newcommand{\Ree}{\operatorname{Re}}
\newcommand{\Imm}{\operatorname{Im}}

\newcommand{\vect}[1]{\bm{#1}}

\definecolor{vs}{rgb}{0.1,0.4,0.1}                  
\definecolor{remcyan}{rgb}{0.0,0.50,0.62}           
\definecolor{corrcyan}{rgb}{0.0,0.60,0.80}          
\ifdefined\cleanversion
  \newcommand{\del}[1]{}                           

  \newcommand{\rem}[1]{}                           
  \newcommand{\delc}[1]{}                          
\else
  \newcommand{\del}[1]{\textcolor{vs}{\sout{#1}}}

  \newcommand{\rem}[1]{\textcolor{remcyan}{[Remark: #1]}}
  \newcommand{\delc}[1]{\textcolor{corrcyan}{\sout{#1}}} 
\fi

\begin{document}

\title{Surface-pinned nonsingular topologically charged defects in nanoplasmonics}

\author{Michael I. Tribelsky}
\email{mitribelsky@gmail.com}
\affiliation{M. V. Lomonosov Moscow State University, Moscow, 119991, Russia}
\affiliation{Moscow Center for Advanced Studies, Kulakova St.\ 20, Moscow, 123592, Russia}

\author{Vladimir V. Ternovski}
\email{vladimir.ternovski@gmail.com}
\affiliation{M. V. Lomonosov Moscow State University, Moscow, 119991, Russia}

\date{\today}

\begin{abstract}
We discuss surface-pinned, topologically charged, nonsingular defects (SDs) in light scattering by a plasmonic object. They follow from the exact Maxwell boundary conditions and are purely topological---unaffected by surface forces. At optical frequencies, in (quasi) 2D, they appear in the electric and Poynting vector fields for $H^{(z)}$ polarization, not for $E^{(z)}$. Exact solutions for a circular cylinder reveal topological charge-conserving bifurcations involving SDs with bulk singularities or SDs alone, and contraction of defect regions under loss.
\end{abstract}

\maketitle 

\emph{Introduction.}
The topology of electromagnetic fields is of broad current interest~\cite{Ni2021,Wang2024,Yang2025,Yan2024}. In resonant nanoparticle scattering, field singularities organize near-field energy flow and local optical response~\cite{Wang2004,Bashevoy2005,Lukyanchuk2006,Tribelsky2006,Lukyanchuk2007}. Conventional defects are isolated field zeros; their local indices and global arrangement determine the topology of the surrounding vector field~\cite{Nye1974,Soskin2001,Dennis2009}.

Here we show that conventional zeros do not exhaust the topology at a sharp material interface. A boundary point acquires an index when a continuous field component has a simple zero there and the other component(s), perpendicular to it, has (have) nonzero one-sided traces of opposite sign. The field therefore does not vanish on each side of the core, yet the defect carries an integer winding number, derived below in closed form. We call such an entity a surface defect (SD).

Boundary-pinned defects, often called boojums, occur in many ordered media, where surface forces commonly affect their structure~\cite{Mermin1979,Volovik1983,Blaauwgeers2002,Volovik2003,Takeuchi2006,Tasinkevych2012}. Here the topological charge follows directly from the Maxwell boundary conditions and the one-sided bulk fields, without introducing a microscopic surface-energy model. Moreover, smoothing the sharp interface into a fictitious transition layer of finite thickness (a purely topological regularization) converts the SDs into conventional singularities; see below.

Somewhat similar defects are discussed in the context of the Kibble–Lazarides–Shafi cosmic walls bounded by strings. However, they acquire a topological charge only in a special regime---near the second of two successive symmetry-breaking transitions, where the two length scales are well
separated~\cite{Volovik2020}. In contrast, the SDs are generic and have a very simple structure (see below).

Because the field is nonzero at the core, these defects permit local surface enhancement rather than enforcing the vanishing of the field, as conventional singularities do. Related streamline patterns were reported previously~\cite{Bashevoy2005,LukyanchukTernovsky2006}, but their interface indices and role in global topological charge accounting were not identified. Meanwhile, pinned to the boundary, the SDs affect the field in their vicinity  more strongly than conventional bulk singularities do. As for the net charge accounting, it becomes misleading without the SDs.

Though the SDs {may} exist in both 3D and 2D problems, for simplicity we consider them only in 2D. The 2D case corresponds to scattering by an infinite cylinder (not necessarily circular) and to patterns in a plane of mirror symmetry for a 3D scatterer. Generalization to essentially 3D cases may be performed with the help of the same approach. However, we must emphasize that in 3D, the SDs may be realized as
both point and line defects.


We show that {in 2D} at optical frequencies for a plasmonic scatterer the SDs occur in $\vect{E}$ and $\vect{S}$ patterns for the $H^{(z)}$ polarization of the incident wave, but do not occur in any $\vect{H}$-field pattern. Also no SDs occur in any of the three fields for the $E^{(z)}$ polarization. This explains the well-known polarization dependence of the field patterns. Here and below $\vect{E}$ and $\vect{H}$ denote the complex electric and magnetic field amplitudes; $\vect{S}$ stands for the time-averaged Poynting vector; the $z$-axis is aligned normal to the  mirror-symmetry plane.

\emph{ Problem formulation.} 
We use Gaussian units and normalize $\vect{E}$, $\vect{H}$, and $\vect{S}$ by their incident-wave amplitudes.
The incident plane, linearly polarized, monochromatic wave propagates in a vacuum ($\mu=\eps=1$). For the cylinder we assume $\mu=1$ (optical range) and $\eps=\eps'+\ii\eps''$ with $\eps'<0$ and $\eps''\geq0$. Here $\eps$ and $\mu$ are the permittivity and permeability of the medium, respectively. For $E^{(z)}$ polarization only $E_z$ and $H_{x,y}$ are nonzero; for $H^{(z)}$ polarization only $H_z$ and $E_{x,y}$ are nonzero. In either case $\vect{S}$ lies in the $xy$ plane.

The Maxwell boundary conditions make $E_\tau$, \mbox{$D_n\equiv\eps E_n$,} and the entire $\vect{H}$ continuous~\cite{Landau2013}. Thus, $\vect{E}$ is continuous for $E^{(z)}$ polarization but not for $H^{(z)}$.

We analyze patterns of $\vect{S}$ and $\vect{\mathcal{E}}\equiv\Ree\vect{E}$. The former does not depend on the initial incident phase $\phi_0$ (the one of the incident wave at the origin of the coordinate system at $t = 0$), but the latter does. A shift of the time origin changes~$\phi_0$. It affects the $\vect{\mathcal{E}}$ pattern, including the positions of defects and their bifurcation points, but does not change the local classes of the SDs (see below). The $\Imm{\vect{E}}$-field patterns obey the same rules as those of $\Ree\vect{E}$.   

Note also that in optics usually only the phase difference between the inspected and reference fields matters. Varying $\phi_0$ of the reference wave by standard methods, one gets the opportunity to control the positions of SDs and bifurcations involving them, which may be important for experimental study of the defects.        

In the lossless case, continuity of $D_n$ makes the one-sided normal components of $\vect{\mathcal{E}}$ opposite in sign. With $D_n=|D_n|\ee^{\ii\alpha_{_{D_n}}}$, reversal persists under loss when   
\mbox{$\eps'+\eps''\tan\alpha_{_{{D_n}}}<0$} (the condition is to be applied away from points where $\Ree D_n=0$). This condition always holds for sufficiently small $\eps''$.

Introduce the complex Poynting vector $\vect{\mathit{\Pi}}$ and the right-handed triad of unit vectors $\vect{n},\,\vect{\tau}$, and $\vect{e}_z$, where $\vect{n}$ is the outward normal to the surface,  $\vect{\tau}$ stands for the tangential vector, and $\vect{e}_z$ is the basis vector of the $z$-axis.  For the normalized fields used here, $\vect{\mathit{\Pi}}\equiv\vect{E}\times\vect{H}^*=\vect{S}+\ii\vect{X}$, where $\vect{X}$ is the reactive part.

For $E^{(z)}$ polarization, $\vect{\mathit{\Pi}}=(-E_zH_\tau^*,E_zH_n^*,0)$ is continuous. No component reverses across the interface.

For $H^{(z)}$ polarization, $\vect{\mathit{\Pi}}=(E_\tau H_z^*,-E_nH_z^*,0)$. Its normal component is continuous, whereas $\mathit{\Pi}^{(\rm in)}_\tau=-P/\eps$ and $\mathit{\Pi}^{(\rm out)}_\tau=-P$, with continuous $P=D_nH_z^*$. Hence $S_\tau$ reverses across a lossless positive-negative-$\eps$ interface. Under loss, the same criterion as that  derived above for $\vect{\mathcal{E}}$ holds with $\alpha_{_{{D_n}}}\to\alpha_{_{P}}$, and reversal persists for sufficiently weak dissipation.

\begin{figure}[!ht]
  \centering
  \includegraphics[width=\columnwidth]{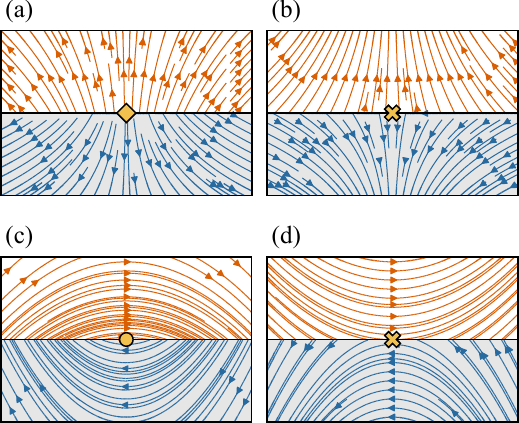}
  \caption{Schematic field lines near an SD: $\Ree\vect{E}$ [(a),(b)] and $\vect{S}$ [(c),(d)]. White and gray colors denote $\eps=1$ and \mbox{$\Ree\eps<0$} respectively. Diamonds/circles mark $w=+1$ (quasi-nodes for  $\Ree\vect{E}$ and quasi-centers/foci for $\vect{S}$); crosses mark $w=-1$ (quasi-saddles for both  $\Ree\vect{E}$ and $\vect{S}$).}\label{fig:SD}
\end{figure}

\emph{  Surface defects and topological charge.}
Let the continuous component $\mathcal{E}_\tau$ or $S_n$ have a simple zero at the boundary while the other component has nonzero one-sided traces.  When those traces have opposite signs, the signs of $\mathcal{E}_n\partial_\tau\mathcal{E}_\tau$ or $S_\tau\partial_\tau S_n$ distinguish the four topologically different local configurations presented in Fig.~\ref{fig:SD}. Here $\partial_\tau$ denotes the oriented tangential derivative. These four configurations exhaust the set of topologically inequivalent  SDs. Because {for our dimensionless fields} $S_n=\Ree(E_\tau H_z^*)$, the electric- and Poynting-field SDs generally do not coincide.
   
\begin{figure}
  \centering
  \includegraphics[width=\columnwidth]{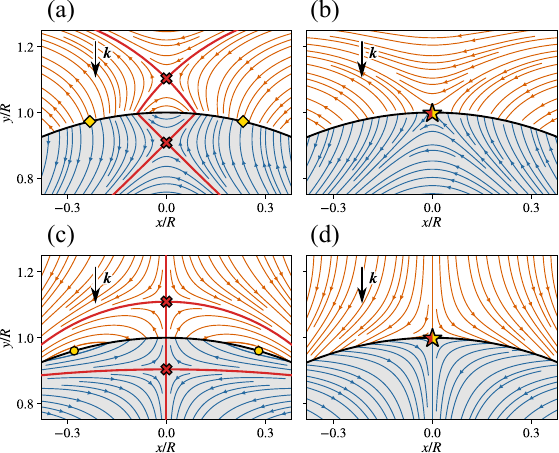}
  \caption{Charge-conserving annihilation involving bulk saddles (crosses) with SDs in exact Mie solution for scattering of \mbox{$H^{(z)}$-polarized} wave by lossless cylinder; normal incidence; {$\phi_0=0$;} $kR=0.3$. The arrow shows the incident wave vector $\vect{k}$; the shaded region is the cylinder. The cylinder axis passes through the origin of the coordinate system. (a) For $\Ree\vect{E}$ at $\eps\approx-1.009$. Two bulk saddles ($w=-1$) and two SDs (quasi-nodes with $w=+1$) give zero total charge. (b) At $\eps\approx-1.008$ they merge into a $w=0$ state (star). (c) For $\vect{S}$ at $\eps\approx-0.982$, two bulk saddles and two SDs (quasi-centers with $w=+1$) are present. (d) They merge at $\eps\approx-0.973$. Beyond the bifurcations the displayed regions are defect-free.}\label{fig:bif}
\end{figure}

\begin{figure}
  \centering
  \includegraphics[width=\columnwidth]{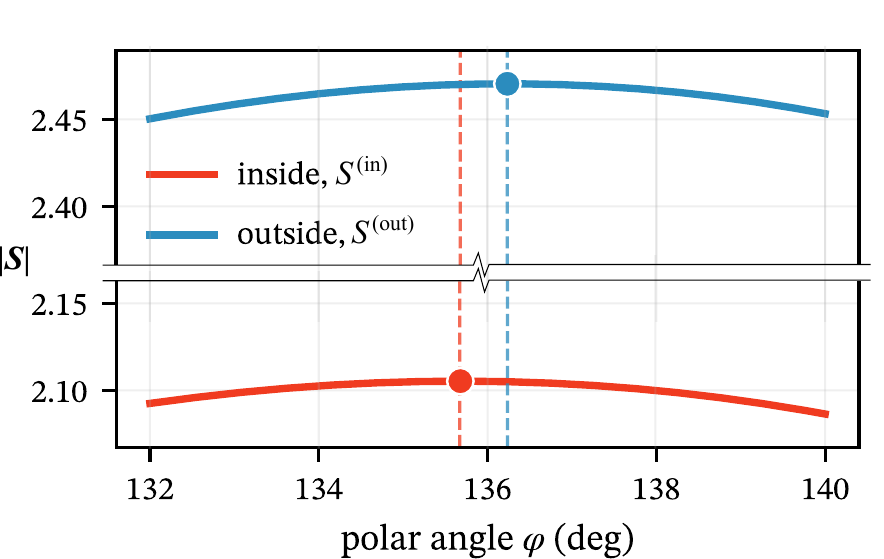}
  \caption{
  Distinct one-sided maxima of $|\vect{S}|$ near an SD of an aluminum nanowire. Exact Mie solution: $R=50$~nm, \mbox{$\lambda=121.25$~nm,} and \mbox{$\eps=-1.144+0.098\ii$.} The angular separation $0.558^\circ$ corresponds to 0.487~nm.}\label{fig:Smax}
\end{figure}

\begin{figure*}
  \centering
  \includegraphics[width=\textwidth]{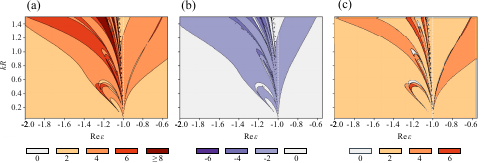}
  \caption{Surface-defect winding-number maps for the Poynting field in scattering of $H^{(z)}$-polarized wave by a nanowire. Exact Mie solution. 
  $\Imm\eps=0$. 
  (a) $W_{\rm SD}^{+}$, (b) $W_{\rm SD}^{-}$, (c) $W_{\rm SD}=W_{\rm SD}^{+}+W_{\rm SD}^{-}$. Because $|w|=1$ for the SD, $|W_{\rm SD}^{\pm}|$ counts the number of defects of each sign. All multipole resonances accumulate at $\eps'=-1$ (dashed line) as $kR\to0$. 
  See text and Supplemental Material~2~\cite{fig1footnote}.}
  \label{fig:Diss}
\end{figure*}

\emph{ Winding number.} 
The counterclockwise winding is $+1$ for Fig.~\ref{fig:SD}(a),(c) and $-1$ for Fig.~\ref{fig:SD}(b),(d). Locally,
\begin{align}
w_{_E} &= -\left.\sgn\left(\mathcal{E}_n\frac{\partial \mathcal{E}_\tau}
{\partial\tau}\right)\right|^{\rm in}_{\rm SD} \equiv
\left.\sgn\left(\mathcal{E}_n\frac{\partial \mathcal{E}_\tau}
{\partial\tau}\right)\right|^{\rm out}_{\rm SD}, \label{eq:wE} \\
w_{_S} &= \left.\sgn\left(S_\tau\frac{\partial S_n}{\partial\tau}
\right)\right|^{\rm in}_{\rm SD} \equiv
-\left.\sgn\left(S_\tau\frac{\partial S_n}{\partial\tau}
\right)\right|^{\rm out}_{\rm SD}. \label{eq:wS} 
\end{align}
Here ``in'' and ``out'' denote one-sided traces evaluated at the SD. Without reversal, $w=0$.

To define $w$ rigorously, we replace the sharp boundary by a finite-width layer with a differentiable interpolation between its edges and use the conventional definition of the winding number:
\begin{equation}\label{eq:w}
  w = \frac{1}{2\pi}\oint\limits_\Gamma d\theta \equiv \frac{1}{2\pi}\oint\limits_\Gamma \nabla\theta\!\cdot\! d\vect{r},
\end{equation}
where contour $\Gamma$ encloses the SD and $\theta$ is the vector angle. We evaluate the contour integral piecewise and, at each oriented interface crossing $A\to B$, choose the angle jump by continuity from neighboring points with $|\Delta\theta|\leq\pi$. Since
$\int_A^B d\theta=\theta(B)-\theta(A)$, this contribution depends only on the one-sided endpoint directions, not on an interpolation across the interface. 
Then we reduce the layer thickness to zero~\footnote{Since the continuous component vanishes at the SD, the regularization stretches the SD into a line, transforming it into a conventional singularity at the intersection of the two null lines. We emphasize that the regularization is purely topological—the fields inside the layer need not satisfy the Maxwell equations.
}. Together with the smooth parts of $\Gamma$, these jumps give Eqs.~\eqref{eq:wE} and \eqref{eq:wS}. Thus, though an SD is not a zero of either one-sided field, it has the same local index as the corresponding conventional singularity. 

\emph{ Global topological charge balance.}
The scattered field vanishes at infinity, and the $\vect{S}$ pattern tends to that of the incident wave, i.e., to a constant. The winding number for this field equals zero. It means that a contour enclosing the scatterer and all defects gives
$W_{\rm B}+W_{\rm SD}=0$. Here $W_{\rm B,\,SD}$ are the net winding indices for the bulk singularities and SDs respectively.  The equality may be considered as a generalized version of the Poincar\'{e}–Hopf index-sum theorem, so the SDs complete the global index accounting.  


\emph{ Nanowire examples.}
Using the precision-controlled code~\cite{ternovski2026arXiv} for visualization, we evaluate the exact Mie solution for an infinite right circular cylinder of radius $R$ under normal incidence. Figure~\ref{fig:bif} shows the visualization results illustrating the general boundary-condition mechanism and the exchange of indices between SDs and bulk singularities. Bifurcations involving SDs alone are presented in Supplemental Material~1~\cite{fig1footnote}. Unlike a conventional singularity, an SD has nonzero one-sided $\vect{S}$; distinct inner and outer local maxima can therefore occur nearby; see Fig.~\ref{fig:Smax}. 

\emph{ Effects of dissipation.}
{Increasing dissipation simplifies the topological structure by removing both singular and nonsingular defects. To illustrate this claim, we present {for the $\vect{S}$-field} a map of domains with different values of the SD winding numbers in the $(\eps', kR)$ plane at $\eps'' = 0$; see Fig.~\ref{fig:Diss}. The map is rather complex. As $\eps''$ increases, domains with large values of the winding numbers disappear one by one, while the shape of the remaining domains smooths out. Eventually, a single nontrivial domain with $W_{\rm SD} = 2$ remains. It has the shape of a narrow wedge, whose lower boundary coincides with the \mbox{$\eps'$-axis,} and whose vertex points in the positive direction of this axis. The rest of the map corresponds to $W_{\rm SD} = 0$. A further increase in $\eps''$ pushes the wedge to the left, out of the region depicted in Fig.~\ref{fig:Diss}, and none of the SDs remains. Details of this evolution are shown in a movie presented in Supplemental Material 2~\cite{fig1footnote}.
}



\emph{Possible experimental evidence.} These defects play an important role in near-field pattern formation. Indirect evidence can be obtained by reconstructing patterns from near-field measurements of electromagnetic interactions with material objects. Direct evidence requires measuring the fields themselves. Though this is a challenging task at the nanoscale, recent advances~\cite{Wu2025,Mokhun2026} give hope that the  task is feasible. Another way is to model the effects at radio frequencies~\cite{Kapitanova2017}.

\emph{Conclusions.}
We have identified  nonsingular boundary-pinned defects carrying interface indices \mbox{$w=\pm1$.} {For plasmonic scatterers in} 2D cases, the Maxwell boundary conditions generate them in $\vect{E}$ and $\vect{S}$ for $H^{(z)}$ (but not $E^{(z)}$) polarization, when a continuous component has a simple zero and the other has opposite-sign, nonzero one-sided traces. Exact solutions for a plasmonic circular cylinder reveal topological charge-conserving bifurcations involving either SDs with bulk singularities or SDs alone. Dissipative losses simplify the field pattern, decreasing the number of defects and contracting the SD-supporting parameter regions (note, however, that these processes may depend on $\eps''$ in a nonmonotonic manner). The same conclusions apply to any multi-component real vector field with this interface structure.

\emph{Data availability.} The numerical data and code that support the findings of this article are available from the corresponding author upon reasonable request.

\begin{acknowledgments}
This work was supported by the state assignment of Lomonosov Moscow State University. The numerical calculations were supported by the Russian Science Foundation under Grant No.~23-72-00037.
\end{acknowledgments}

\bibliography{Tern_votices}
\end{document}